# "E. A. Milne and the Universes of Newton and Relativistic Cosmology"

by

J. Dunning-Davies,
Department of Physics,
University of Hull,
Hull HU6 7RX,
England.

J.Dunning-Davies@hull.ac.uk

**Abstract.**

Here the 1930's work of Milne on the relationship between the universes of relativistic cosmology and those which follow from Newtonian theory is reviewed. The extension to the case of non-zero pressure is considered also. In each case, any assumptions made are noted and the thermodynamic implications of these are explored in the final section.

# 1. Introduction.

In the 1930's, Milne[1] initiated an investigation into the relationship between the universes of relativistic cosmology and those which may be considered using only Newtonian theory. In this work, he was later joined by McCrea[2] and himself gathered together all the results in his book on relativity, gravitation and world structure [3]. It seems somewhat surprising that this work does not appear well-known these days. One reason for this may be that Milne and McCrea concentrated on the zero pressure situation. Here it is intended to review the approach of Milne and McCrea but also to examine the situation obtaining when the pressure is not zero. This latter case has been alluded to by Peebles [4] and Harrison [5]. In addition, Harrison pointed out that Newtonian cosmology provides an excellent description of a universe in which the pressure is small. Further, it might be noted that Callen, Dicke and Peebles [6] claim the Newtonian treatment to be a perfectly correct method. Here more detail will be presented to enable the assumptions, particularly the thermodynamic assumptions, made to become more obvious and for the implications of these to be discussed.

# 2. A résumé of Milne's approach.

Following an earlier paper by Milne in which the velocity $v$ was assumed equal to the escape velocity [1], McCrea and Milne [2] investigated the case when the velocity does not necessarily have this value. $v$ was assumed to be the velocity of a particle at a distance $r$ from the observer at time $t$. This velocity was assumed radial in nature and a function of $r$ and $t$. Under these circumstances, the equation of motion

$$\frac{D\boldsymbol{v}}{Dt} = \boldsymbol{F},$$

or

$$\frac{\partial \boldsymbol{v}}{\partial t} + \boldsymbol{v} \cdot \frac{\partial \boldsymbol{v}}{\partial r} = \boldsymbol{F} \qquad (2.1)$$

where $\boldsymbol{F}$ is the force due to gravity and is given by Poisson's equation

$$\nabla \cdot \boldsymbol{F} = -4\pi G\rho,$$

where $\rho$ is a function of $t$ only.

The equation of continuity may be written

$$\frac{1}{\rho}\frac{d\rho}{dt} + \frac{1}{r^2}\frac{\partial}{\partial r}(r^2 v) = 0.$$

It follows that the second term in this equation must be a function of $t$ only. Hence, put

$$\frac{1}{\rho}\frac{d\rho}{dt} = -3f(t),$$

so that

$$\frac{1}{r^2}\frac{\partial}{\partial r}(r^2 v) = 3f(t),$$

which may be integrated to give

$$r^2 v = r^3 f(t) + g(t),$$

where $g(t)$ is a constant of integration. This may be rewritten

$$v = rf(t) + r^{-2}g(t) \quad (2.2)$$

If this expression is inserted into (2.1), then, since $\rho$ is a function of $t$ only,

$$\frac{1}{r}[rf'(t) + r^{-2}g'(t) + \{rf(t) + r^{-2}g(t)\}\{f(t) - 2r^{-3}g(t)\}]$$

must be a function only of $t$ also. Hence, $g(t)=0$ and (2.2) becomes

$$v = rf(t) \quad (2.3)$$

which may be integrated to give

$$r = \alpha R(t),$$

where $\alpha$ is a constant arising from the integration and $R(t)$ is a function of $t$ satisfying

$$\frac{1}{R}\frac{dR}{dt} = f(t) = -\frac{1}{3\rho}\frac{d\rho}{dt}. \quad (2.4)$$

Hence,

$$\rho = \beta/R^3, \quad (2.5)$$

where $\beta$ is a constant.

Taking the divergence of (2.1), substituting for $v$ from (2.3) and using Poisson's equation gives

$$3\left\{\frac{df}{dt} + f^2\right\} = -4\pi G\rho \quad (2.6)$$

Using (2.4) and (2.5) in (2.6) then leads to

$$\frac{1}{R}\frac{d^2R}{dt^2} = -\frac{4\pi G\beta}{3R^3},$$

which may be integrated to give

$$\left(\frac{dR}{dt}\right)^2 = \frac{8\pi G\beta}{3R} - \gamma,$$

where $\gamma$ is a constant. By using (2.5), these latter two equations may be written

$$\frac{1}{R}\frac{d^2R}{dt^2} = -\frac{4\pi G\rho}{3} \quad (2.7)$$

and

$$\left(\frac{1}{R}\frac{dR}{dt}\right)^2 + \frac{\gamma}{R^2} = \frac{8\pi G\rho}{3} \quad (2.8)$$

respectively. These final two equations are seen to be formally identical with the equations usually associated with relativistic cosmology for an expanding universe with zero pressure.[4]

The above is the quite standard procedure adopted by McCrea and Milne and which appears also in Milne's book [3]. However, it is worth noting that it applies solely to the case of zero pressure and, frequently these days, the equations used in relativistic cosmology involve pressure.

### 3. A modification of the above to involve pressure.

If the pressure is to be taken into account, equation (2.1) would become

$$\frac{D\mathbf{v}}{Dt} + \frac{1}{\rho}\nabla p = \mathbf{F},$$

where $p$ is the pressure and $\mathbf{F}$ the force due to gravity. However, if, as is usually assumed, the pressure is a function only of time $t$, then $\nabla p$ will be zero. The pressure will actually enter the problem via the modified Poisson equation applicable in this case. As has been pointed out by Peebles, a generalisation to the case where the pressure is not negligible is provided by noting that the source for gravity changes from the mass density $\rho$ to $(\rho + 3p)$. Here an equilibrium between matter and radiation is being assumed and the $3p$ term is taking account of the radiation pressure. Thus, for this case, Poisson's equation becomes

$$\nabla \cdot \mathbf{F} = -4\pi G(\rho + 3p).$$

The only effect of this modification would be to alter (2.7) to

$$\frac{1}{R}\frac{d^2 R}{dt^2} = -\frac{4}{3}\pi G(\rho + 3p) \qquad (3.1)$$

Again following Peebles, it might be noted that, since $\rho$ is the mass per unit volume, the net energy within the sphere is $U = \rho V$. When the material moves so that the radius of the sphere changes, the energy $U$ changes also because of the work due to pressure on the surface:

$$dU = -p\, dV = \rho\, dV + V\, d\rho$$

Hence,

$$\dot{\rho} = -(\rho + p)\dot{V}/V = -3(\rho + p)\dot{R}/R$$

where the volume $V \propto R^3$. Substituting for $p$ in (3.1) gives

$$\ddot{R} = -\frac{4\pi G}{3}\left\{\rho - \frac{R\dot{\rho}}{\dot{R}} - 3\rho\right\}R = \frac{8\pi G \rho}{3}R + \frac{4\pi G}{3}\frac{R^2 \dot{\rho}}{\dot{R}},$$

where the dot refers to differentiation with respect to $t$. This latter equation may be integrated to give

$$\dot{R}^2 = \frac{8\pi G}{3}\rho R^2 + const.$$

Following the approach of McCrea and Milne once more, this latter equation is seen to be of exactly the same form as (2.8). Hence, a slight modification of the approach of McCrea and Milne is seen to lead to equations of the same form as those of relativistic cosmology in the case of a non-zero pressure.

### 4. Discussion of assumptions and results.

In the above, it is immediately obvious that the final equations, derived by utilising purely Newtonian methods are identical in form with those resulting from the more modern relativistic techniques. In section 3, dealing with the case of non-zero pressure, it is instructive to look more closely at the various assumptions made. In noting that the energy $U$ changes because of work due to pressure, the equation

$$dU = -pdV$$

emerges. This is a special case of

$$TdS = dU + pdV - \mu dN,$$

which applies, for example, when both $TdS$ and $\mu dN$ equal zero; that is, when the process under consideration is adiabatic and the total number of particles remains unaltered. Hence, once again it is seen that the equations used for cosmological discussions imply adiabaticity. This was the conclusion reached by looking directly at the Einstein equations used to introduce the idea of inflation and, at that time, it was pointed out that these equations could not be used to describe non-adiabatic situations. [7]. Here it is made clear that the equations apply also only to situations in which particle number is conserved. The importance of these observations lies in the fact that they place very clear limitations on the use of the said Einstein equations and on the equations of identical form derived by Newtonian methods.

It might be noted, however, that an alternative explanation for the use of the equation

$$dU = -pdV$$

in place of

$$TdS = dU + pdV - \mu dN$$

does exist; that is that any entropy change is brought about purely by a change in particle number. This, and this alone, would enable non-adiabaticity to be allowable within the models and here the plural is used because this argument applies equally well to the use of the equations of both Newton and Einstein.

Further, while the methods employed here to derive these basic equations differ greatly from those normally used in general relativity and the meanings of some of the symbols may vary, the two sets of equations are formally identical and the situations they are supposed to describe are the same. Therefore, the question concerning the place and importance of the accepted equations of general relativity must be raised. In their paper, McCrea and Milne, having derived the equations in the case of zero pressure, go on to discuss the curvature of space and make the point that 'the local properties of the universes in expanding spaces of positive, zero or negative curvatures are observationally the same as in Newtonian universes with velocities respectively less than, equal to, or greater than the parabolic velocity of escape.' This is further claimed to give 'great insight into the physical significance of expanding curved space.' It is of immediate interest to note that they always talk of 'space' not 'space-time', thus keeping the three-dimensional world in which we live and time as two separate concepts firmly at the forefront of any considerations. This has the effect of making it immediately clear what is claimed to be happening in our surroundings.